\documentclass[journal,twoside,web]{ieeecolor}
\usepackage{generic}
\usepackage{cite}
\usepackage{amsmath,amssymb,amsfonts}
\usepackage{algorithmic}
\usepackage{graphicx}
\usepackage{textcomp}
\usepackage{amsmath,amsfonts}
\usepackage{algorithmic}
\usepackage{algorithm}
\usepackage{array}
\usepackage[caption=false,font=normalsize,labelfont=sf,textfont=sf]{subfig}
\usepackage{textcomp}
\usepackage{stfloats}
\usepackage{url}
\usepackage{verbatim}
\usepackage{graphicx}
\usepackage{cite}
\usepackage{booktabs}
\usepackage{subfig}
\usepackage{float}
\usepackage{nomencl}
\usepackage[table]{xcolor}
\usepackage{multirow}
\usepackage{makecell}
\usepackage{balance}
\usepackage{verbatim}
\usepackage[bottom]{footmisc}
\def\BibTeX{{\rm B\kern-.05em{\sc i\kern-.025em b}\kern-.08em
    T\kern-.1667em\lower.7ex\hbox{E}\kern-.125emX}}
\begin{document}
\title{Grid-Constrained State-Adaptive Particle Swarm Optimization: A Discrete and Efficient Heuristic Solver for Precise Harmonic Programming}
\vspace{-0mm}
\author{Guangze Chen, Zhenbin Zhang, Yafei Yin
\vspace{-11mm}
}

\maketitle

\begin{abstract}
Harmonic programmed pulse width modulation (HPPWM), offering flexible harmonic regulation, is a promising solution for high-power energy conversion systems. However, most existing methods solve HPPWM in a continuous space while ignoring the finite timer resolution of practical digital controllers. This leads to a potential optimality deviation during deployment. Motivated by this, this paper proposes a \textit{Grid-Constrained State-Adaptive Particle Swarm Optimization (GCSA-PSO)} strategy. By matching the solution space with practical timer constraints, GCSA-PSO directly searches for implementable pulse sequences in the discrete solution space, thereby improving deployment consistency while reducing the search burden. Moreover, a state-adaptive evaluation strategy is developed to assign different cost evaluations according to particle states, avoiding unnecessary evaluations and improving computational efficiency. Experimental data confirm that, compared with the classical method, the proposed method reduces the computational time while achieving higher control accuracy under practical digital-controller deployment.
\end{abstract}


\section{Introduction}
\label{sec:introduction}
\IEEEPARstart{A}{s} traditional fossil fuels decline and environmental challenges intensify, wind power, particularly high-power wind energy installations, has emerged as one of the key solutions to energy issues~\cite{blaabjerg2013future}. In order to reduce the considerable power loss during energy conversion, high-power converters typically operate at low switching frequencies, which inevitably degrades output power quality. 

Harmonic programmed pulse width modulation (HPPWM), owing to its exceptional capability to optimize harmonic distributions, has become a promising solution~\cite{7112491}. Based on Fourier theory, HPPWM establishes an explicit relationship between switching pulses and harmonic components, through which the desired switching angles can be obtained by solving the corresponding optimization problem or algebraic equations. This enables flexible allocation of the harmonic spectrum. Owing to its flexibility, various HPPWM variants have been developed for different objectives, including selective harmonic elimination~\cite{6541241,9369141,9551787}, total harmonic distortion~\cite{11190002,11216011,8902012}, dc-link voltage ripple~\cite{11424991}, torque ripple~\cite{11421049,10982199,10214477}, grid codes~\cite{11419864,4376284,11397218}. However, the nonlinear solving process introduced by Fourier-based formulation remains a major challenge, making the determination of HPPWM switching angles quite difficult~\cite{10877753}.

%Harmonic programmed pulse width modulation (HPPWM), owing to its exceptional capability to optimize harmonic distributions, has become a promising solution~\cite{7112491}. Based on Fourier theory, HPPWM establishes an explicit relationship between switching pulses and harmonic components, through which the desired switching angles can be obtained by solving the corresponding optimization problem or algebraic equations. This enables flexible allocation of the harmonic spectrum. Owing to its flexibility, various HPPWM variants have been developed for different objectives, including selective harmonic elimination~\cite{6541241,9369141,9551787}, total harmonic distortion~\cite{11190002,11216011,8902012}, dc-link voltage ripple~\cite{11424991}, torque ripple~\cite{11421049,10982199,10214477}, grid codes~\cite{11419864,4376284,11397218}. However, the nonlinear solving process introduced by Fourier-based formulation remains a major challenge, making the determination of HPPWM switching angles quite difficult~\cite{10877753}.

To address the above challenges, substantial research efforts have been devoted to the efficient determination of HPPWM switching angles over the past few decades. From a methodological perspective, existing approaches can generally be classified into three categories: 1) numerical gradient-based methods, 2) algebraic-based methods, and 3) intelligence-based methods.

Numerical gradient-based methods, as the earliest solving branch, have been utilized since the birth of HPPWM~\cite{4158397}. The basic principle of this class of methods is to update the candidate solution iteratively along the direction determined by mathematical gradients until convergence is achieved. Benefiting from the well-defined search direction, these methods usually exhibit fast convergence~\cite{5612683}. However, their performance is highly dependent on the initial values, i.e., once the initial value is chosen, the final solution is determined. Their lesser capability to escape local optima remains a major limitation~\cite{MEMON20182235}.

Algebraic-based methods provide another important route for solving HPPWM by reformulating the original trigonometric equations into a purely algebraic problem~\cite{10294200}. Typically, these methods transform the trigonometric equations into polynomial forms through Chebyshev theory, and then determine the analytical solutions using dedicated algebraic techniques. Owing to their rigorous mathematical foundation, algebraic-based methods are able to identify feasible solutions of specific HPPWM formulations with high accuracy. However, these methods are fundamentally designed for solving equation systems rather than optimization problems, making them difficult to apply to multi-objective scenarios such as THD optimization.

With the rapid advancement in processors, intelligent-based solving methods have gained significant traction~\cite{1525002,taghizadeh2010harmonic,9366352,9695347,kavousi2011application,kumar2019selective,routray2019harmonic,8946532}. By constructing flexible cost functions, these methods can incorporate different optimization objectives into the HPPWM formulation and provide enhanced global search capability compared with conventional gradient-based methods. Representative examples include genetic algorithms~\cite{1525002}, particle swarm optimization~\cite{taghizadeh2010harmonic,9366352,9695347}, bee algorithm~\cite{kavousi2011application}, whale optimization~\cite{kumar2019selective}, gray wolf optimization~\cite{routray2019harmonic}, and gravitational search algorithm~\cite{8946532}, colonial competitive algorithm~\cite{7038159}, etc. These methods have shown strong applicability in solving nonlinear HPPWM problems, especially when multiple objectives or complex constraints are involved~\cite{yigit2023comparison}.

\begin{figure*}[!t]
	\centering
	\includegraphics[height=2.5in]{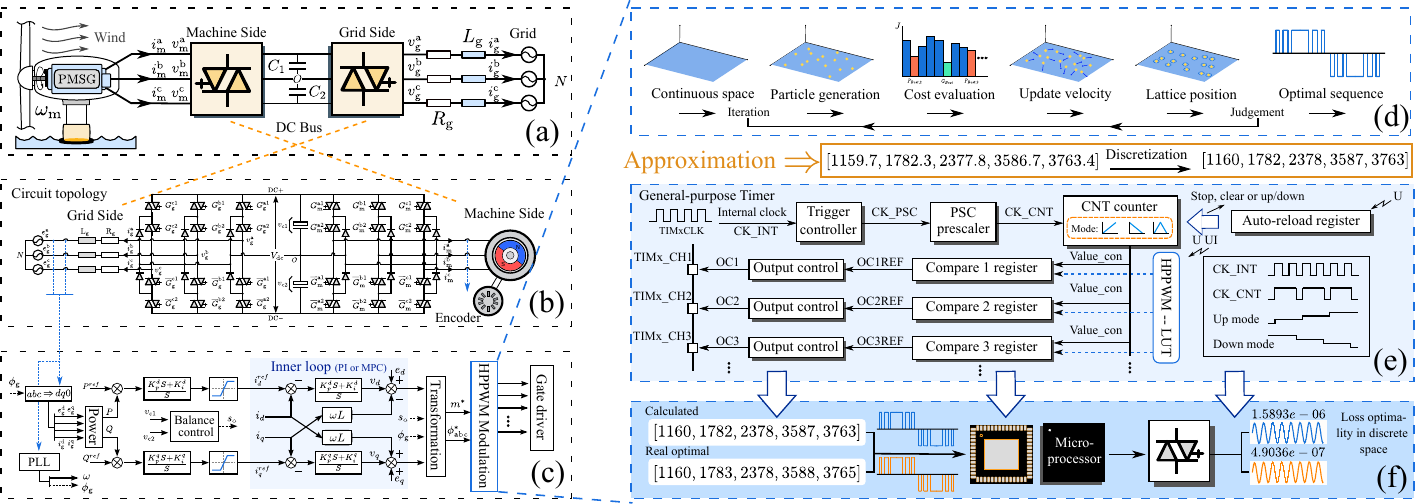}
	\caption{Deployment diagram of HPPWM under high-power energy conversion systems. (a) High-power offshore wind energy generation systems. (b) Topology for power circuits. (c) Typical control scheme. (d) Calculation of HPPWM. (e) Diagram of a general timer used to generate a HPPWM (STM32F103). (f) Optimality deterioration under discrete transformation.}
	\label{fig_1}
    \vspace{-0mm}
\end{figure*}

Building on these intelligent optimization methods, hybrid schemes have been further developed to improve the trade-off between global exploration and local convergence~\cite{shen2014elimination,memon2021asynchronous,padmanaban2021artificial}. A representative example is the hybrid PSO--Newton-Raphson (PSO--NR) method~\cite{shen2014elimination}, where PSO is first used to locate a promising solution region, and NR is then applied as a local refinement step to improve the convergence accuracy. Similarly, the asynchronous particle swarm optimization--genetic algorithm (APSO--GA) combines asynchronous particle evolution with genetic operations to enhance population diversity and convergence performance~\cite{memon2021asynchronous}.

 %Due to the discrete characteristic of real-time digital signal processors, the ideal pulse sequence obtaining in the continuous space requires an approximation which can make it lose the optimality (see, Fig.1).
 
The aforementioned three categories and related hybrid schemes have improved HPPWM solving from different perspectives and have achieved satisfactory performance. Nevertheless, these methods are generally formulated in an assumed continuous solution space, whereas practical digital controllers can only generate switching instants with finite timer resolution. As shown in Fig.~\ref{fig_1}, the obtained continuous switching angles must therefore be discretized before deployment. During this process, the rounded pulse sequence may not strictly coincide with the optimum in the practical discrete solution space, which can introduce an optimality deviation under digital implementation.

Motivated by this, a Grid-Constrained State-Adaptive Particle Swarm Optimization (GCSA-PSO) method is proposed in this work. By aligning the solution space with practical discrete constraints, the proposed method can directly search for the optimum in the discrete space, thereby preserving the optimality in practice while reducing the search space. In addition, a state-adaptive cost evaluation strategy is designed, in which different cost evaluations are assigned to particles under different constraint conditions. This avoids unnecessary computations and further improves the efficiency and computation speed. The major advantages include:
\begin{enumerate}
    \item \textit{Direct discrete solving capability:} GCSA-PSO directly solves HPPWM in the practical discrete solution space, thereby avoiding the post-discretization process required by conventional continuous-space methods and improving consistency with practical digital deployment.
    \item \textit{Reduced search load:} By restricting the candidate switching angles to practical timer-grid points, the grid-constrained solution space excludes infeasible regions that cannot be realized under practical digital-controller constraints, thereby reducing the search domain.
    %By discretizing the solution space, it reduces the particle swarm's search in unnecessary regions of the original space, enhancing search efficiency and facilitating the attainment of a global optimum; 
    \item \textit{Improved computational efficiency:} By assigning state-adaptive cost evaluations to particles under different states, GCSA-PSO avoids redundant cost evaluations and reduces unnecessary computational burden, leading to shorter solving time.
    %The various evaluating strategies for different individuals prevents the inefficient assessment among the numerous particles. This can greatly improve the evaluation efficiency and reduce the computational time.
\end{enumerate}

The rest is structured as follows. Sec.~II describes the basic principles of HPPWM. Sec.~III introduces the proposed GCSA-PSO solving algorithm. Experimental verification, conducted under both general computational test benches and hardware-in-the-loop platforms, is provided in Sec.~IV. Finally, Sec.~V concludes the paper.

\section{Construction of Harmonic Programmed Pulse Width Modulation}

\subsection{Basic Principle}

Over the past few decades, harmonic programmed modulation has been widely developed to improve power quality under low-switching-frequency operation. Although the optimized targets of HPPWM can be diverse, a basic root can be traced to construct the whole theory, i.e. Fourier series decomposition, as given by:
\begin{equation}
\label{equ1}
F_{\rm N }\left(t\right)=\frac{a_0}{2}+\sum_{n=1}^{\rm N }\left(a_n\cos\left(\frac{2\pi nt}{T}\right)+b_n\sin\left(\frac{2\pi nt}{T}\right)\right),
\end{equation}
where \text{$F_{\rm N }$} describes the pulse sequence and \text{$a_{0}$}, \text{$a_{n}$}, \text{$b_{n}$} represent the coefficients of DC-, cosine-, sin-component, respectively. Through mathematical manipulation of Eq.~\eqref{equ1}, the coefficients can be expressed as
\begin{subequations}\label{equ2}
  \begin{align}
    a_n & =\frac{2}{T}\int_{t_0}^{t_0+T}F_{\rm N}(t)\cos(n\omega t)dt\\
    b_n & =\frac{2}{T}\int_{t_0}^{t_0+T}F_{\rm N}(t)\sin(n\omega t)dt.
  \end{align}
\end{subequations}

\begin{figure}[!t]
	\centering
	\includegraphics[width=3.4in]{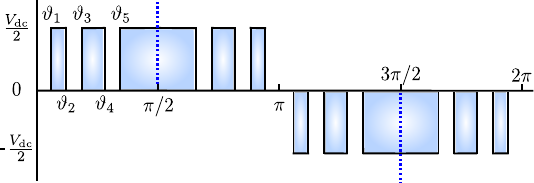}
	\caption{Three-level \& quarter and half wave symmetric pulse sequence (N=5).}
	\label{fig_2}
\end{figure}

Take HPPWM with the target of harmonic elimination as a case study (refer to S-HPPWM in the following). The case waveform is shown in Fig.~\ref{fig_2}, which has three-level and quarter and half wave symmetry(QaHWS). Due to the QaHWS of the pulse sequence, the dc component, cosine terms, and even-order harmonics are naturally eliminated. Therefore, the fundamental and harmonic voltages of the case pulse sequence can be expressed as
\begin{equation}
\label{equ3}
\left\{
\begin{array}{lr}
\frac{2U_{\mathrm{dc}}}{\pi}\sum\limits_{i=1}^{\mathrm{N}}(-1)^{i+1}\cos(\vartheta_i) = V_{\mathrm{basic}} \\
\frac{2U_{\mathrm{dc}}}{n_1\pi}\sum\limits_{i=1}^{\mathrm{N}}(-1)^{i+1}\cos(n_1\vartheta_i)=V_{\mathrm{h},1} &  \\
\ \ \ \ \ \ \ \ \ \ \ \ \ \ \   \vdots\\
\frac{2U_{\mathrm{dc}}}{n_{\mathrm{N}-1}\pi}\sum\limits_{i=1}^{\mathrm{N}}(-1)^{i+1}\cos(n_{\mathrm{N}-1}\vartheta_i)=V_{\mathrm{h},\mathrm{N}-1}, & 
\end{array}
\right.
\end{equation}
where $\rm N$ represents the number of switching angles under QaHWS, \text{$n$} denotes the harmonic order, $V_{\mathrm{h}}$ denotes the harmonic voltage of the pulse sequence, which is expected to be eliminated, and \text{$\vartheta_{i}$} denotes the switching angle. For ease of notation, define \text{$\mathrm{MI}$} as the equivalent modulation index, 
\begin{equation}
\label{equx}
\mathrm{MI}=\frac{\pi V_{\mathrm{basic}}}{2U_{\mathrm{dc}}},
\end{equation}
where $V_{\mathrm{basic}}$ is the amplitude of the fundamental voltage component, and $U_{\mathrm{dc}}$ is the dc-link voltage.

Due to the nature of time series shown in Fig.~\ref{fig_2}, the switching angles needs to meet the following constraints:
\begin{equation}
\label{equ4}
\underbrace{0<\vartheta_{1}<\vartheta_{2}<\vartheta_{3}<\ldots<\vartheta_{\mathrm{N}-1}<\vartheta_{\mathrm{N}}<\frac{\pi}{2}}_{\rm sequence \ constraint}.
\end{equation}

\textit{Example:} For clarity, a three-level QaHWS S-HPPWM with three degrees of freedom, i.e., three switching instants within a quarter period, is considered. In this case, two selected harmonic orders can be eliminated while the desired fundamental component is maintained. Therefore, the corresponding equation set can be established as
\begin{equation}
\label{equ5}
\begin{aligned}
&\left\{
\begin{array}{lr}
\cos(\vartheta_1)-\cos(\vartheta_2)+\cos(\vartheta_3) = \mathrm{MI} \\
\frac{2U_{\mathrm{dc}}}{3\pi}\left[\cos(3\vartheta_1)-\cos(3\vartheta_2)+\cos(3\vartheta_3)\right]=0 &  \\
\frac{2U_{\mathrm{dc}}}{5\pi}\left[\cos(5\vartheta_1)-\cos(5\vartheta_2)+\cos(5\vartheta_3)\right]=0. &  \\
\end{array}
\right. \\
& \ \ \ \ \ \ \ \ \ \ \ \ \ \ 0<\vartheta_{1}<\vartheta_{2}<\vartheta_{3}<\frac{\pi}{2}
\end{aligned}
\end{equation}

\subsection{Optimization Problem}

To enhance the flexibility of HPPWM, the original equation-solving problem is often reformulated as an optimization problem, so that different control objectives and practical constraints can be incorporated in a unified manner~\cite{10877753}. For this purpose, the nonlinear equations can be converted into a comprehensive cost function through simple mathematical manipulation, as
\begin{equation}
\label{equ6}
\left\{
\begin{array}{lr}
\delta_{0}=\sum\limits_{i=1}^{\mathrm{N}}(-1)^{i+1}\cos(\vartheta_i) - \mathrm{MI} \\
\delta_{1}=\frac{1}{n_1}\sum\limits_{i=1}^{\mathrm{N}}(-1)^{i+1}\cos(n_1\vartheta_i) &  \\
\ \ \ \ \ \ \ \ \ \ \ \ \ \ \   \vdots\\
\delta_{N-1}= \frac{1}{n_{N-1}}\sum\limits_{i=1}^{\mathrm{N}}(-1)^{i+1}\cos(n_{N-1}\vartheta_i), &  
\end{array}
\right.
\end{equation}
\begin{equation}
\label{equ7}
J= \underbrace{\gamma_{\mathrm{f}} \times \delta_{0}^{2}}_{\rm fund.} +  \underbrace{\gamma_{\mathrm{h}} \times (\delta_{1}^{2} + \delta_{2}^{2} + \ldots + \delta_{\mathrm{N}-1}^{2})}_{\rm harmonic},
\end{equation}
where $\delta$ represents the cost terms derived from the original nonlinear equations, and $\gamma_{\mathrm{f}}$ and $\gamma_{\mathrm{h}}$ denote the weighting coefficients for the fundamental and harmonic components, respectively. In addition, the sequence constraint in Eq.~\eqref{equ4} should be considered to ensure the practical feasibility of the obtained switching angles.

\section{The Proposed GCSA-PSO Algorithm}

To obtain switching angles with the desired harmonic characteristics, the optimization problem formulated in Eq.~\eqref{equ7} needs to be solved. Existing HPPWM solving methods usually operate in a continuous solution space, whereas practical digital controllers can only generate switching instants with finite timer resolution. Therefore, continuously optimized switching angles need to be discretized before implementation. This discretization may affect the relative optimality of the obtained switching angles and influence the harmonic performance under practical deployment, as illustrated in Fig.~\ref{fig_1}.

Meanwhile, conventional intelligent optimization methods usually evaluate all iterative individuals using the same cost function. For HPPWM solving, however, the contribution of different particles to the search process is not identical. Particles that violate the sequence constraints are already infeasible, and their cost values are mainly determined by the constraint item. Similarly, particles that remain unchanged during the iteration do not provide new cost information. Therefore, applying the full harmonic cost evaluation to all particles may introduce unnecessary trigonometric calculations. This observation motivates the design of a state-adaptive cost evaluation strategy to improve computational efficiency while maintaining the effectiveness of the search process.

Motivated by the above, a Grid-Constrained State-Adaptive Particle Swarm Optimization (GCSA-PSO) strategy is proposed in this section. In the proposed GCSA-PSO framework, each particle represents a candidate HPPWM pulse sequence and is described by its position $\boldsymbol{x}$ and velocity $\boldsymbol{v}$. The position corresponds to the switching angles in the pulse sequence, i.e., $\boldsymbol{x}=\boldsymbol{\vartheta}=[\vartheta_1,\vartheta_2,\dots,\vartheta_N]$, while the velocity determines the update direction and step size during the iterative search. For example, a particle with $\boldsymbol{x}=[15,30,55]$ and $\boldsymbol{v}=[1,2,5]$ represents the current switching-angle solution $[\vartheta_1,\vartheta_2,\vartheta_3]=[15,30,55]$, which is updated to $[16,32,60]$ in the next step. 

\begin{figure}[!t]
	\centering
	\hspace{+2mm}
	\includegraphics[width=2.9in]{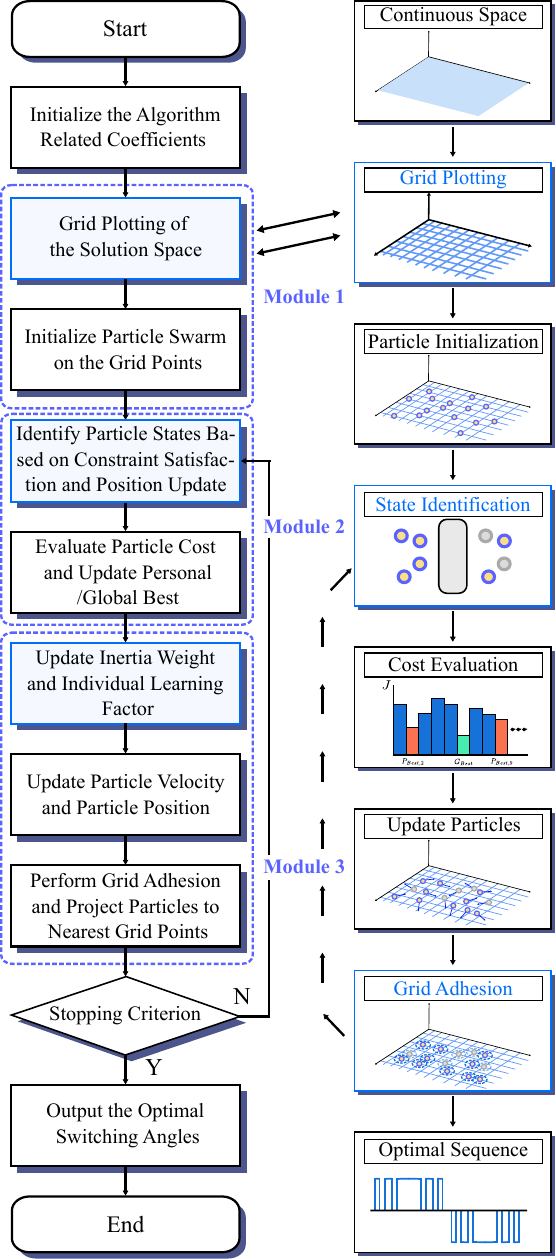}
	\caption{Overall diagram of the proposed GCSA-PSO algorithm.}
	\label{fig_4}
	\vspace{-0mm}
\end{figure}

As shown in Fig.~\ref{fig_4}, the overall procedure of GCSA-PSO consists of three main modules corresponding to the following subsections. \textit{1) Grid plotting and initialization:} the finite timer resolution is mapped into the solution space, and the initial particles are generated directly on realizable grid points. \textit{2) State-adaptive cost evaluation:} the full cost function and constraint item are formulated, and particles are classified into different states according to their update status and constraint satisfaction, so that state-specific cost evaluations can be assigned. \textit{3) Update and reposition:} the personal best and global best solutions are updated, followed by the velocity-position update and grid adhesion, which projects the updated particles back onto practical grid points.

The detailed implementation of each step is introduced in the following subsections, and the overall pseudo-code is provided in Algorithm~\ref{GCSA-PSO}.

\subsection{Grid Plotting and Initialization}

To embed the finite timer resolution of practical digital controllers into the solution space, grid plotting is introduced so that GCSA-PSO directly searches over implementable switching angles. The corresponding grid interval is determined by the timer basic clock and the fundamental frequency as
\begin{equation}
\label{equ8}
d_{\rm grid}=\frac{f_{\rm b}}{f_{\rm p}} \times \xi_{\rm cv} \times 360,
\end{equation}
where $f_{\rm p}$ represents the basic timer clock frequency, $f_{\rm b}$ is the fundamental frequency of the desired pulse sequence, and $\xi_{\rm cv}$ denotes the conversion coefficient for the computational unit of the solution space. For example, $\xi_{\rm cv}=1$ when the solution space is expressed in degrees, while $\xi_{\rm cv}=\pi/180$ when it is expressed in radians.

Through this grid plotting process, the proposed method initializes particles directly on grid points and searches for the optimum in the practical discrete solution space. This allows the obtained switching angles $\boldsymbol{\vartheta}$ to preserve its optimality under practical digital-controller deployment.

\subsection{State-Adaptive Cost Evaluation Strategy}

Considering the priority of fundamental tracking, harmonic suppression, and sequence constraints, the full evaluation function is formulated as
\begin{equation}
\label{equ9}
J_{\rm full}
=
\gamma_{\mathrm{f}}\times\delta_{0}^{2}
+
\gamma_{\mathrm{h}}\times\sum_{j=1}^{N-1}\delta_{j}^{2}
+
\gamma_{\mathrm{c}}\times\delta_{c},
\end{equation}
where $\delta_0$ represents the fundamental-amplitude deviation, $\delta_j$ denotes the deviation of the selected $j$th harmonic component. The constraint item $\delta_c$ is introduced to penalize pulse sequences that violate the switching-angle order in~\eqref{equ4}, which is expressed as
\begin{equation}
\label{equ10}
\delta_{c}
=
\left[
u(-\vartheta_1)
+
\sum_{i=1}^{N-1}u(\vartheta_i-\vartheta_{i+1})
+
u\left(\vartheta_N-\frac{\pi}{2}\right)
\right],
\end{equation}
where $u(\cdot)$ is the step function used to represent inequality violations. Therefore, $\delta_c=0$ indicates that the particle satisfies the sequence constraint in Eq.~\eqref{equ4}, whereas $\delta_c>0$ indicates an infeasible pulse sequence.

In conventional algorithms, all particles are usually evaluated using the same full cost function. For HPPWM solving, however, the contribution of different particles to the search process is not identical. Particles satisfying the sequence constraints provide valid harmonic information and therefore require full evaluation. In contrast, particles violating the sequence constraints are already infeasible, and their cost values are mainly determined by the constraint item. For these particles, evaluating the harmonic terms introduces additional trigonometric calculations but contributes little to the final judgment. Similarly, particles that remain unchanged during the iteration can directly inherit their previous cost values. Based on this observation, a state-adaptive cost evaluation strategy is proposed.

\begin{algorithm}[!t]
\caption{Proposed GCSA-PSO Solving Algorithm}
\label{GCSA-PSO}
\renewcommand{\algorithmicrequire}{\textbf{Input:}}
\renewcommand{\algorithmicensure}{\textbf{Output:}}
\begin{algorithmic}[1]
	\REQUIRE $\omega_{\rm max}$, $\omega_{\rm min}$, $c_{1,\rm max}$, $c_2$, $P$, $v_{\rm max}$, $t_{\rm max}$, $\mathrm{MI}$, $f_{\rm p}$, $f_{\rm b}$, etc.
	\ENSURE $\boldsymbol{\vartheta}=[\vartheta_1,\vartheta_2,\dots,\vartheta_N]$.
	\STATE Calculate the grid interval $d_{\rm grid}$ according to Eq.~\eqref{equ8}.
	\FOR{each particle $i=1,2,\dots,P$}
		\STATE Initialize $\boldsymbol{x}_{i}^{0}$ on the grid points and initialize $\boldsymbol{v}_{i}^{0}$.
		\STATE Set $\boldsymbol{P}_{\rm Best,i}^{0}=\boldsymbol{x}_{i}^{0}$ and evaluate $J_i^{0}$.
	\ENDFOR
	\STATE Set $\boldsymbol{G}_{\rm Best}^{0}$ as the particle with the minimum cost value.
	\FOR{$t=1$ to $t_{\rm max}$}
		\FOR{each particle $i=1,2,\dots,P$}
			\STATE Calculate the constraint item $\delta_c$ according to Eq.~\eqref{equ10}.
			\IF {$\|\boldsymbol{v}_{i}^{t-1}\|_{\infty}<d_{\rm grid}/2$}
				\STATE Set $S_i=2$ \hfill // virtual state
			\ELSIF {$\delta_c>0$}
				\STATE Set $S_i=1$ \hfill // semi-virtual state
			\ELSE
				\STATE Set $S_i=0$ \hfill // effective state
			\ENDIF

			\IF {$S_i=0$}
				\STATE Evaluate the full cost $J_i^{t}=J_{\rm full}$ according to Eq.~\eqref{equ9}.
			\ELSIF {$S_i=1$}
				\STATE Evaluate the simplified cost $J_i^{t}=\gamma_{\mathrm{c}}\times\delta_{c}$.
			\ELSE
				\STATE Inherit the cost $J_i^{t}=J_i^{t-1}$.
			\ENDIF

			\STATE Update $\boldsymbol{P}_{\rm Best,i}^{t}$ and $\boldsymbol{G}_{\rm Best}^{t}$ according to $J_i^{t}$.
			\STATE Update $\omega^{t}$ and $c_1^{t}$ according to Eq.~\eqref{equ14} and Eq.~\eqref{equ15}.
			\STATE Update the velocity $\boldsymbol{v}_{i}^{t}$ according to Eq.~\eqref{equ11}.
			\STATE Limit each element of $\boldsymbol{v}_{i}^{t}$ within $[v_{\rm min},v_{\rm max}]$.
			\STATE Update the position $\boldsymbol{x}_{i}^{t}$ according to Eq.~\eqref{equ12}.
			\STATE Attach $\boldsymbol{x}_{i}^{t}$ to the nearest grid points through the grid adhesion process.
			% \STATE Limit each element of $\boldsymbol{x}_{i}^{t}$ within the boundary of the solution space.
		\ENDFOR
		\IF {the stopping criterion $\varepsilon_{\boldsymbol{x}}$ is satisfied}
			\STATE \textbf{break}
		\ENDIF
	\ENDFOR
	\RETURN $\boldsymbol{\vartheta}=\boldsymbol{G}_{\rm Best}$.
\end{algorithmic}
\end{algorithm}

\begin{figure}[!t]
	\vspace{-2mm}
	\centering
	\includegraphics[width=3in]{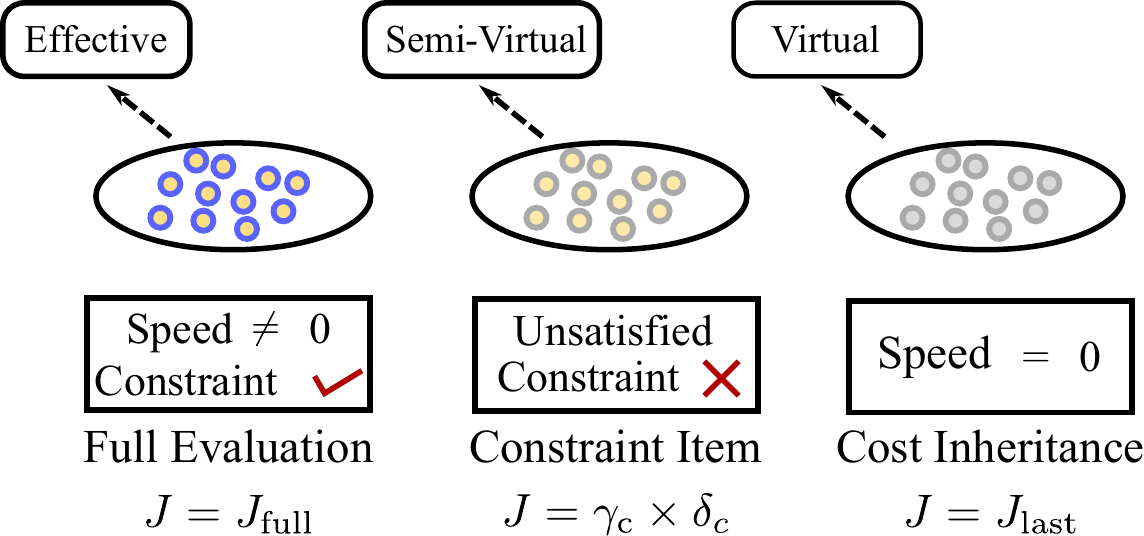}
	\vspace{-0mm}
	\caption{Particle states and the corresponding state-adaptive cost evaluation strategies.}
	\label{fig_5}
\end{figure}

Here, GCSA-PSO classifies particles into three states according to their status and constraint satisfaction, as shown in Fig.~\ref{fig_5}:
\begin{enumerate}
	\item \textit{Effective state:} the particle is updated and satisfies the constraint, i.e., $\|\boldsymbol{v}_{i}^{t-1}\|_{\infty}>d_{\rm grid}/2$ and $\delta_c=0$.
	\item \textit{Semi-virtual state:} the particle is updated but violates the constraint, i.e., $\|\boldsymbol{v}_{i}^{t-1}\|_{\infty}>d_{\rm grid}/2$ and $\delta_c>0$.
	\item \textit{Virtual state:} the particle remains unchanged after grid adhesion, i.e., $\|\boldsymbol{v}_{i}^{t-1}\|_{\infty}<d_{\rm grid}/2$.
\end{enumerate}
For effective particles, the full cost function in Eq.~\eqref{equ9} is evaluated to assess both harmonic performance and constraint satisfaction. For semi-virtual particles, only the constraint item is used, i.e., $J=\gamma_{\mathrm{c}}\times\delta_{c}$, because their infeasibility has already been identified by the sequence constraint. For virtual particles, the cost value is inherited from the previous iteration, i.e., $J_i^{t}=J_i^{t-1}$, since their positions remain unchanged. In this way, the evaluation is assigned according to the particle state, reducing unnecessary trigonometric calculations and improving the computational efficiency of the solving process.

\subsection{Update and Reposition Strategy}

After cost evaluation, the personal best solution and the global best solution are updated. The particle velocity is then updated according to the inertia term, individual learning term, and social learning term, as
\begin{equation}
\label{equ11}
\begin{split}
\boldsymbol{v}_{i}^{t}
=
&\omega^{t}\boldsymbol{v}_{i}^{t-1}
+
c_{1}^{t}r_{1}
\left(
\boldsymbol{P}_{\rm Best,i}^{t}
-
\boldsymbol{x}_{i}^{t-1}
\right)\\
&
+
c_{2}r_{2}
\left(
\boldsymbol{G}_{\rm Best}^{t}
-
\boldsymbol{x}_{i}^{t-1}
\right),
\end{split}
\end{equation}
where $\boldsymbol{x}_i$ and $\boldsymbol{v}_i$ are the position and velocity of the $i$th particle, respectively. $\boldsymbol{P}_{\rm Best,i}$ is the best position historically found by the $i$th particle, and $\boldsymbol{G}_{\rm Best}$ is the best position found by the entire swarm. $r_1$ and $r_2$ are independent random numbers within $[0,1]$, and $t$ denotes the current iteration index. Subsequently, the position is updated as
\begin{equation}
\label{equ12}
\boldsymbol{x}_{i}^{t}
=
\boldsymbol{x}_{i}^{t-1}
+
\boldsymbol{v}_{i}^{t}.
\end{equation}
Since the random terms in the velocity update may drive the updated position away from the predefined discrete grid points, a grid adhesion operation is further applied, i.e.,
\begin{equation}
\label{equ13}
\boldsymbol{x}_{i}^{t}
\leftarrow
d_{\rm grid}\cdot
\mathrm{round}
\left(
\frac{\boldsymbol{x}_{i}^{t}}{d_{\rm grid}}
\right).
\end{equation}
This operation projects each updated particle onto the nearest realizable grid point, ensuring that every candidate pulse sequence remains implementable by the digital controller.

To balance the global exploration capability and convergence speed, the inertia weight is dynamically adjusted as
\begin{equation}
\label{equ14}
\omega^{t}
=
\omega_{\rm max}
-
\frac{\omega_{\rm max}-\omega_{\rm min}}{t_{\rm max}}t.
\end{equation}
Therefore, a relatively large inertia weight is assigned in the early iterations to enhance global exploration, while a smaller inertia weight is achieved in the later iterations to improve convergence.

In addition, the individual learning factor is gradually reduced after a threshold iteration, so as to weaken the influence of local personal-best directions in the later stage:
\begin{equation}
\label{equ15}
c_{1}^{t}
=
\begin{cases}
c_{1,\rm max}, & t<t_{\rm tv},\\
c_{1,\rm max}-\dfrac{c_{1,\rm max}(t-t_{\rm tv})}{t_{\rm max}-t_{\rm tv}}, & t\geq t_{\rm tv}.
\end{cases}
\end{equation}
where $t_{\rm tv}$ is the threshold iteration for learning-factor adjustment. Through the above update and reposition strategy, GCSA-PSO ensures that the switching angles obtained during the search process remain consistent with practical discrete implementation. Meanwhile, the dynamic weighting mechanism achieves a balance between solution-space exploration and convergence speed.

\begin{table}[tbp]
	\caption{Parameters of the Implemented GCSA-PSO}
	\label{table1}
	\centering
	\renewcommand{\arraystretch}{1.08}
	\resizebox{\columnwidth}{!}{
	\begin{tabular}{l l c c}
		\hline\hline
		Symbol & Description & Value & Unit\\
		\hline
		$\gamma_{\mathrm{f}}$ & Weighting of fundamental term & 50 & -\\
		$\gamma_{\mathrm{h}}$ & Weighting of harmonic term & $1$ & -\\
		$\gamma_{\mathrm{c}}$ & Weighting of constraint violation & $10^{5}$ & -\\
		$c_{1,\max}$ & Maximum individual learning factor & 0.5 & -\\
		$c_{2}$ & Social learning factor & 0.5 & -\\
		$P$ & Population size & 300 & -\\
		$\omega_{\min}$ & Minimum inertia weight & 0.4 & -\\
		$\omega_{\max}$ & Maximum inertia weight & 1.5 & -\\
		$t_{\max}$ & Maximum iteration count & 200 & -\\
		$t_{\rm tv}$ & Threshold iteration for $c_1$ adjustment & 100 & -\\
        $\varepsilon_{\boldsymbol{x}}$ & Position convergence tolerance & $10^{-4}$ & -\\
		$v_{\max}$ & Particle velocity upper limit & 3.6 & deg\\
		$v_{\min}$ & Particle velocity lower limit & -3.6 & deg\\
		$f_{\mathrm{p}}$ & Timer frequency & 1 & MHz\\
		$f_{\mathrm{b}}$ & Fundamental frequency & 50 & Hz\\
		$\xi_{\mathrm{cv}}$ & Conversion coefficient & 1 & -\\
		\hline\hline
	\end{tabular}
	}
\end{table}

\section{Performance Validation}

To validate the effectiveness of the proposed GCSA-PSO, a general computational test bench and a hardware-in-the-loop (HIL) platform were constructed in the laboratory. The platform consists of one general computation unit with an Intel i5-14490F CPU and two RT-Box3 produced by PLECS. The general computation unit is used to deploy the solving algorithms and evaluate their computational speed. For the HIL test, one RT-Box3 is used as the digital controller to implement the calculated HPPWM lookup-tables (LUTs), while the other RT-Box3 is used to emulate the power converter. The main parameters of GCSA-PSO are listed in Table~\ref{table1}. For a fair comparison, the common parameters, including population size, maximum iteration number, and initial conditions, are kept identical for the compared methods.

\begin{figure}[!t]
	\centering
	\includegraphics[width=3.2in]{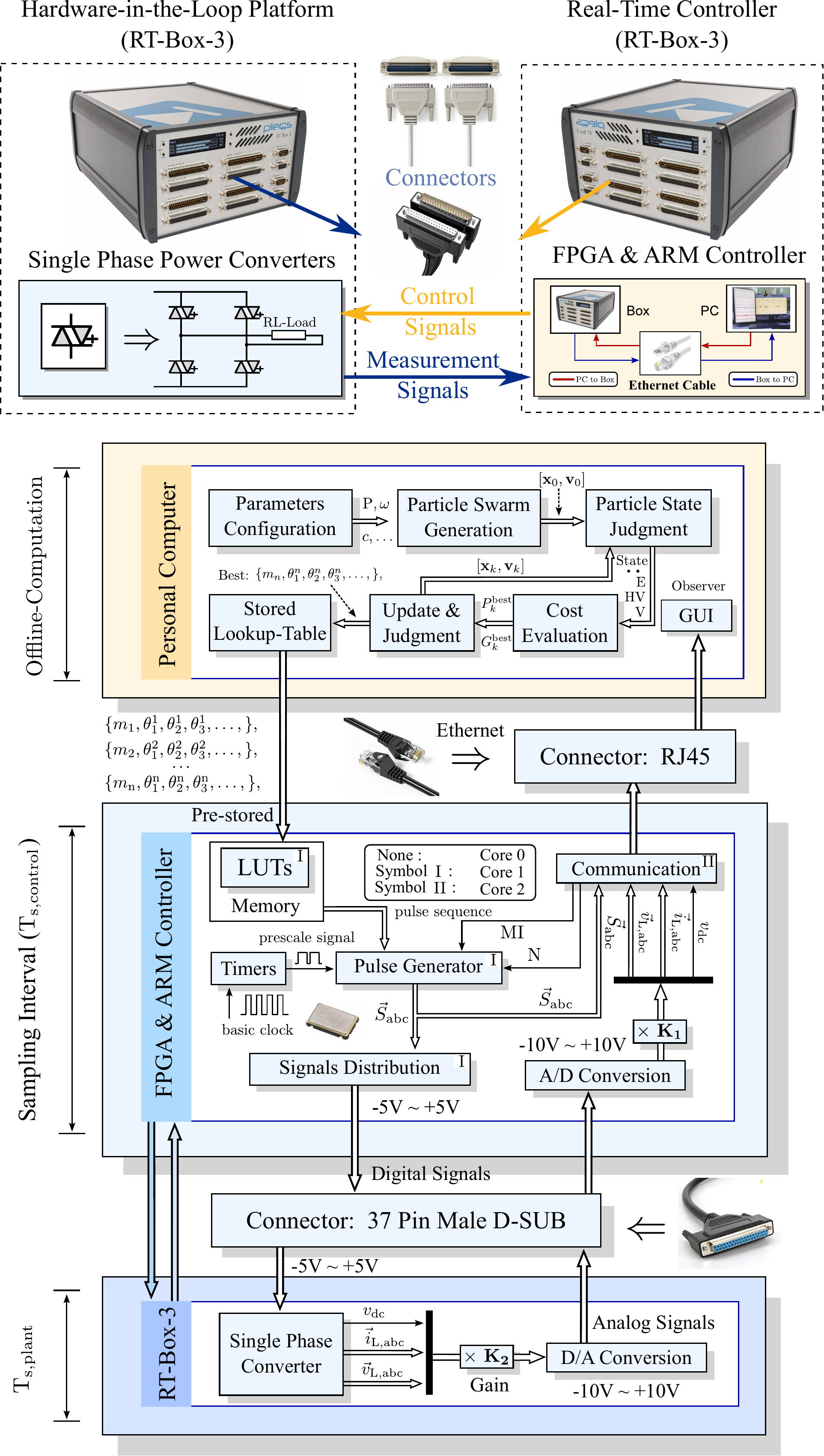}
	\caption{Deployment diagram for the constructed test bench.}
        \vspace{-0mm}
	\label{fig_6}
\end{figure}

\begin{figure*}
	\centering
 \hspace{+10mm}
	\subfloat[]{
		\begin{minipage}[t]{0.33\linewidth}
			\centering
            \hspace{-10mm}
			\includegraphics[width=2.1in]{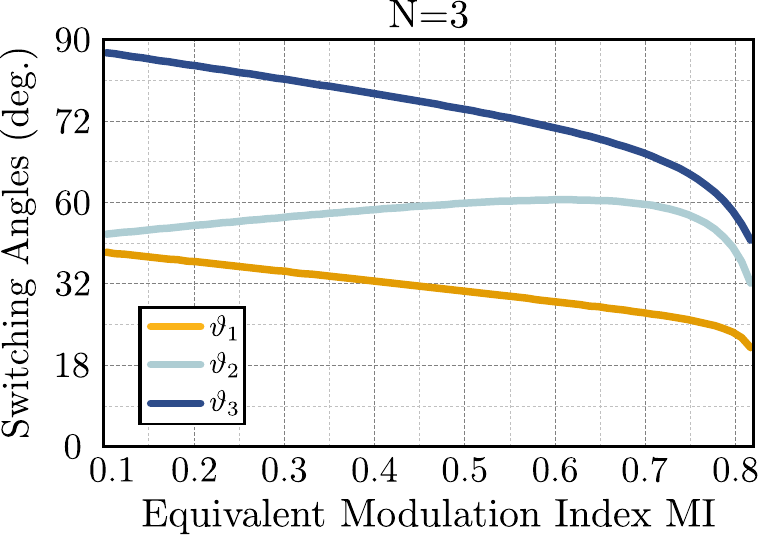}
		\end{minipage}
	}%
	\subfloat[]{
		\begin{minipage}[t]{0.33\linewidth}
			\centering
   \hspace{-10mm}
			\includegraphics[width=2.1in]{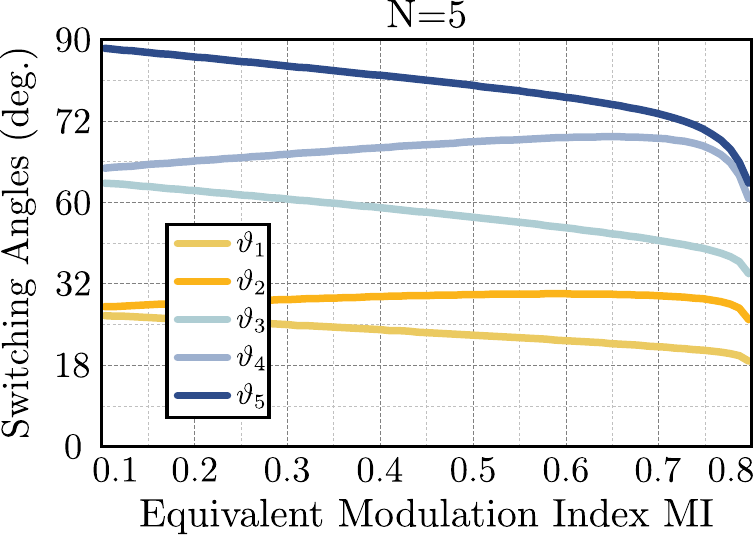}
		\end{minipage}
	}%
 \subfloat[]{
		\begin{minipage}[t]{0.33\linewidth}
			\centering
   \hspace{-10mm}
			\includegraphics[width=2.1in]{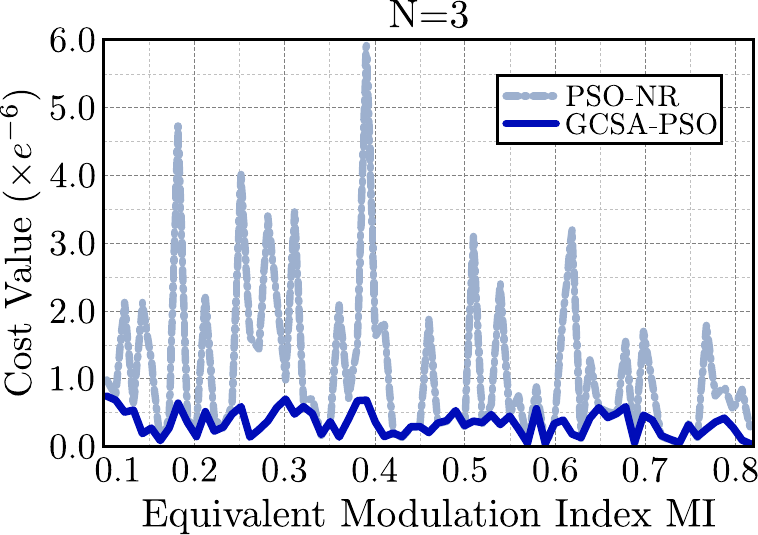}
		\end{minipage}
	}%
 
 \subfloat[]{
		\begin{minipage}[t]{0.33\linewidth}
			\centering
   \hspace{-10mm}
			\includegraphics[width=2.1in]{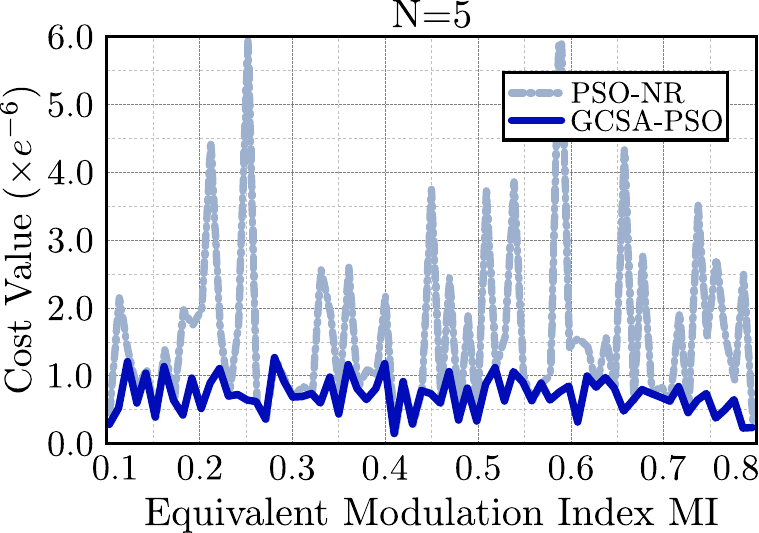}
		\end{minipage}
	}%
	\subfloat[]{
		\begin{minipage}[t]{0.33\linewidth}
			\centering
   \hspace{-10mm}
			\includegraphics[width=2.1in]{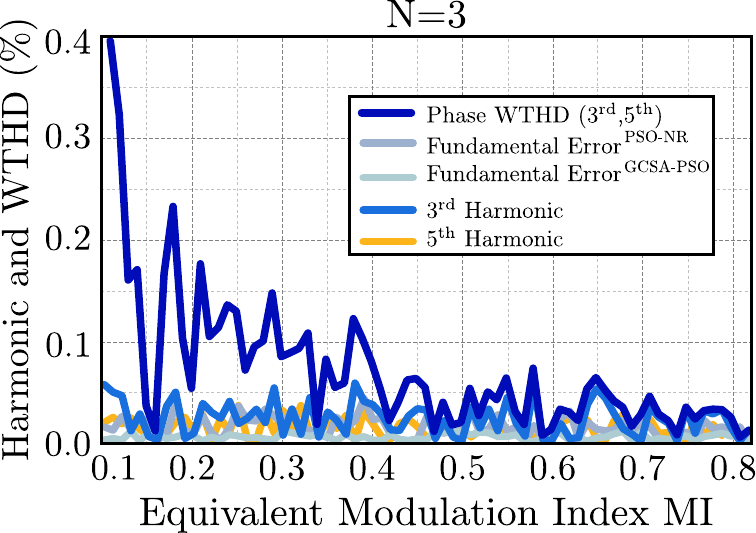}
		\end{minipage}
	}%
 \subfloat[]{
		\begin{minipage}[t]{0.33\linewidth}
			\centering
   \hspace{-10mm}
			\includegraphics[width=2.1in]{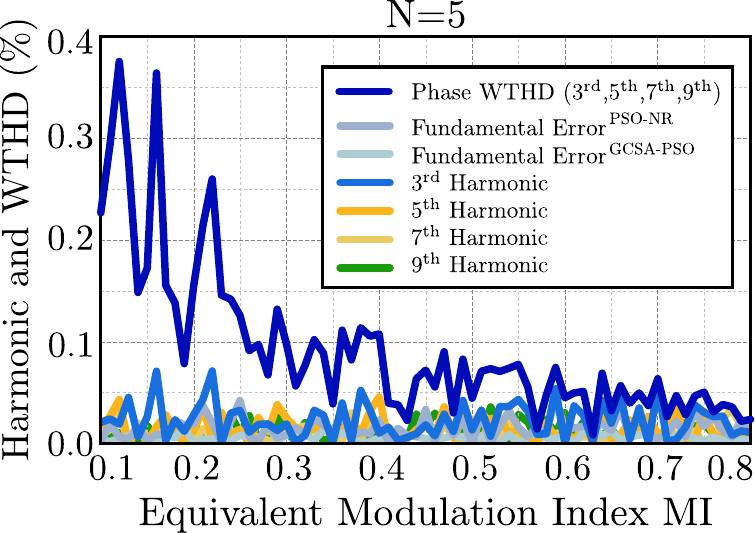}
		\end{minipage}
	}%
	\centering
	\caption{[PC and HIL Results] Optimal discrete pulse sequence and associated performance. (a) Discrete pulse sequence (N=3). (b) Discrete pulse sequence (N=5). (c) Optimal cost value (N=3). (d) Optimal cost value (N=5). (e) Harmonics and WTHD (N=3). (f) Harmonics and WTHD (N=5).}
	\label{fig_7}
    \vspace{-0mm}
\end{figure*}

\begin{figure}
\vspace{-3mm}
	\centering
	\subfloat[\text{$\mathrm{N}=3$}, \text{$\mathrm{MI}=0.53$}.]{
		\begin{minipage}[t]{1\linewidth}
			\centering
			\includegraphics[width=2.85in]{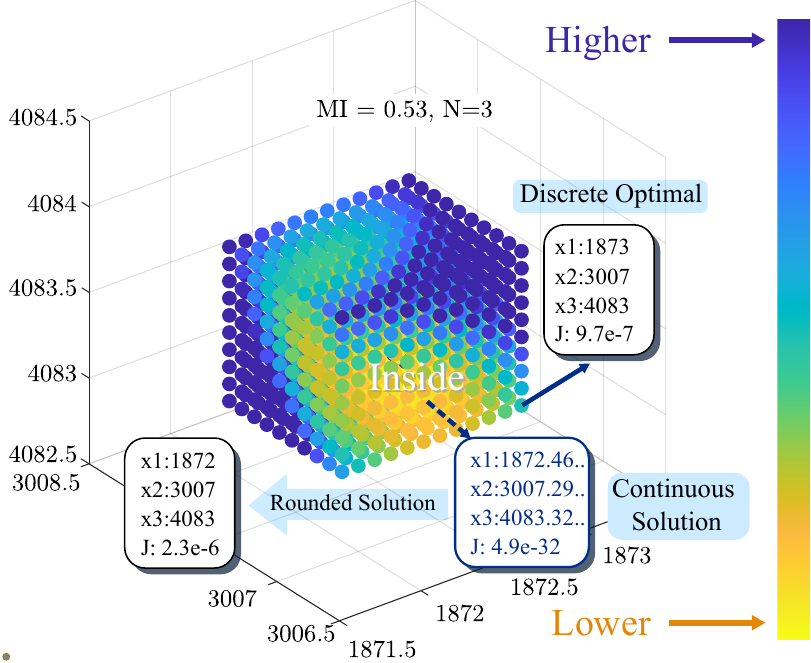}
		\end{minipage}
	}%
 
	\subfloat[\text{$\mathrm{N}=3$}, \text{$\mathrm{MI}=0.79$}.]{
		\begin{minipage}[t]{1\linewidth}
			\centering
			\includegraphics[width=2.85in]{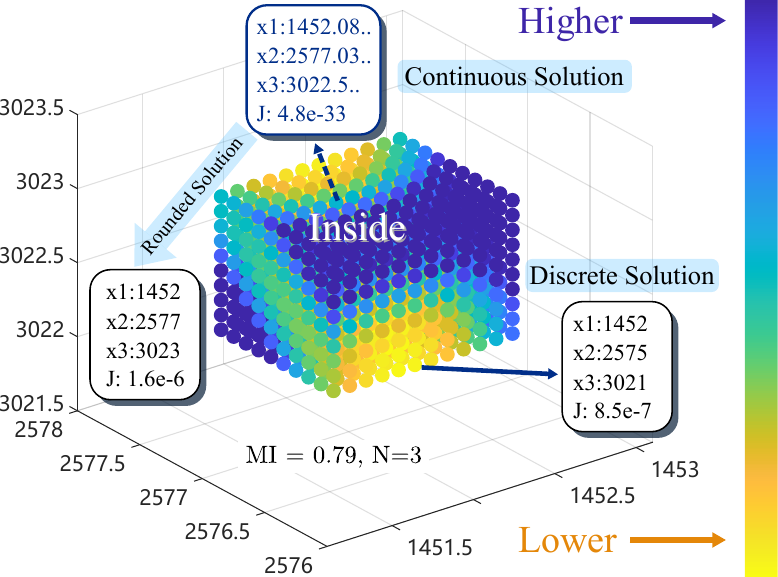}
		\end{minipage}
	}%
	\centering
	\caption{[PC Results] Cost distribution around the continuous-space optimum and the corresponding discrete grid points. (a) \text{$\mathrm{N}=3$}, \text{$\mathrm{MI}=0.53$}. (b) \text{$\mathrm{N}=3$}, \text{$\mathrm{MI}=0.79$}.}
	\label{fig_8}
 \vspace{-0mm}
\end{figure}

\begin{figure}[!t]
	\centering
    \hspace{-5mm}
	\includegraphics[width=3.3in]{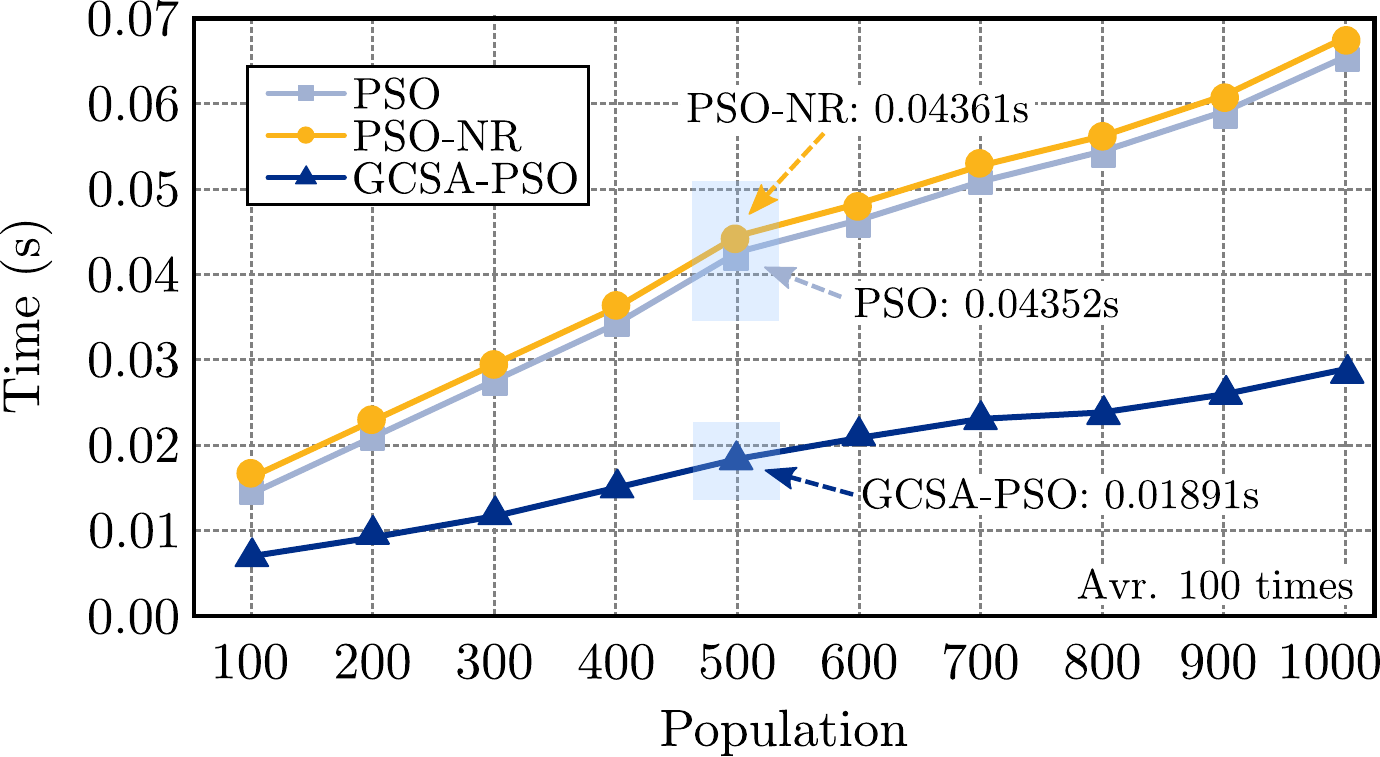}
	\caption{[PC Results] Computational-speed comparison among PSO, PSO-NR, and GCSA-PSO for S-HPPWM with $\mathrm{N}=5$. Each operating condition is tested 100 times, and the averaged results are reported.}
	\label{fig_9}
    \vspace{-0mm}
\end{figure}

\begin{figure*}[!t]
	\centering
	\includegraphics[width=7.1in]{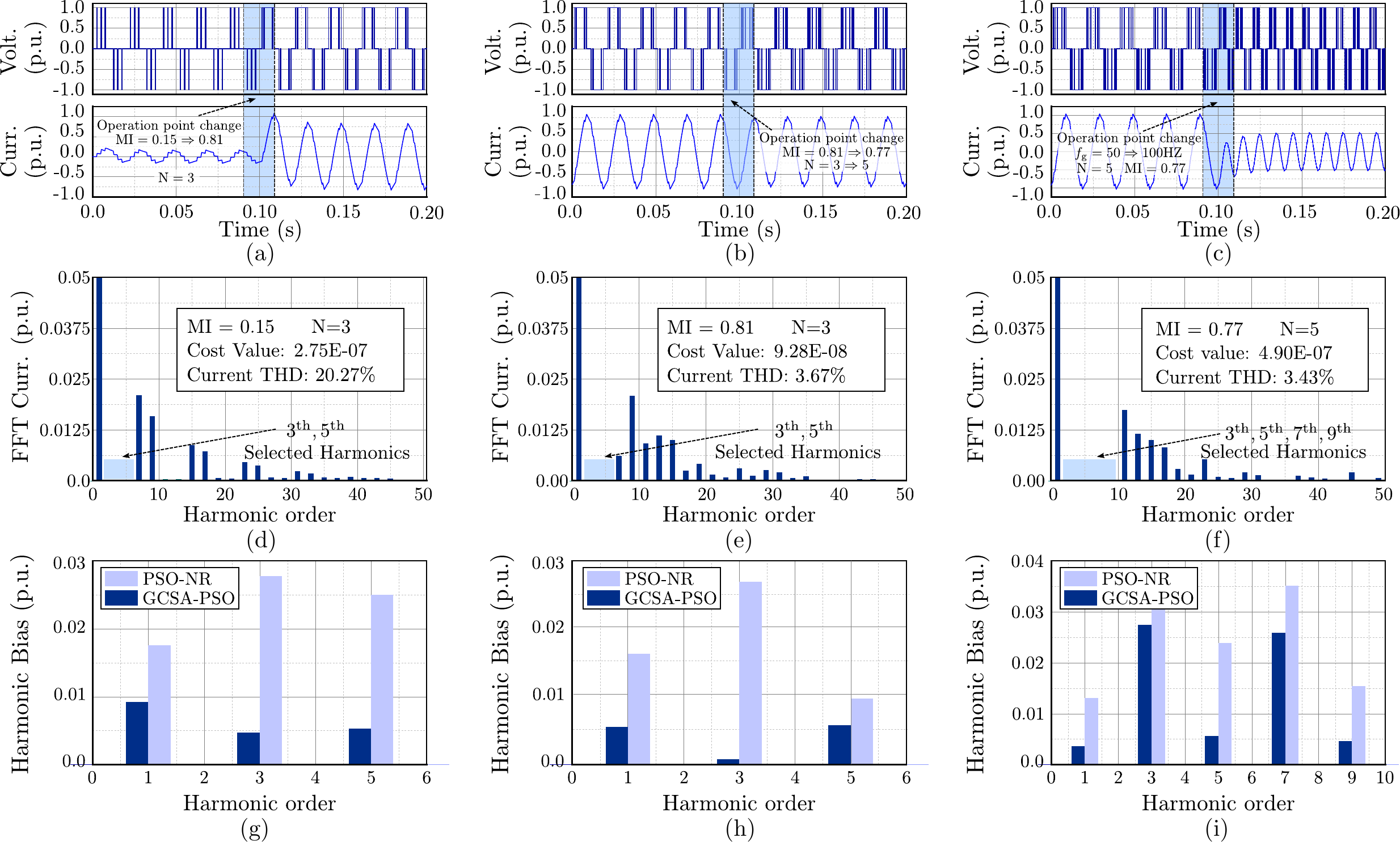}
	\caption{[HIL Results] Performance of HPPWM under various scenarios. (a) Change of MI (0.15 to 0.81). (b) Change of MI (0.81 to 0.77). (c) Change of frequency
		from 50 to 100 Hz. (d) Current FFT (MI=0.34, N=3). (e) Current FFT (MI=0.81, N=3). (f) Current FFT (MI=0.77, N=5). (g) Control bias (MI=0.34, N=3). 
		(h) Control bias (MI=0.81, N=3). (i) Control bias (MI=0.77, N=5).}
  \vspace{-0mm}
	\label{fig_10}
\end{figure*}

The validation is organized into three parts: 1) optimality verification, which demonstrates that the optimum obtained by classical continuous-space methods may become non-optimal after deployment on a practical digital controller, whereas the proposed GCSA-PSO can preserve the discrete optimality; 2) computational-speed test, which compares the solving time of different methods under the same initial conditions, iteration number, and population size; and 3) controller-deployment test, which verifies the harmonic regulation accuracy of the obtained pulse patterns under practical digital-controller implementation. The deployment diagram is illustrated in Fig.~\ref{fig_6}, and the experimental results are shown in Figs.~\ref{fig_7}--\ref{fig_10}. The corresponding analysis is provided in the following.

\subsection{Optimality Verification}

To verify the discrete optimality of the pulse sequences solved by the proposed method, a comparative test was conducted between GCSA-PSO and PSO-NR. In this test, S-HPPWM cases with three and five degrees of freedom were calculated and the basic timer clock of the digital controller is set at 1~MHz. The results are shown in Fig.~\ref{fig_7}, where Figs.~\ref{fig_7}(a) and (b) present the solved switching angles under different modulation indices for $N=3$ and $N=5$, respectively. Figs.~\ref{fig_7}(c) and (d) show the corresponding cost values $J_{\mathrm{best}}$ of the pulse sequences under practical discrete implementation conditions. It can be observed that GCSA-PSO consistently reaches the discrete global optimum, whereas the solutions obtained by PSO-NR exhibit optimality deterioration after discretization. This is because GCSA-PSO embeds the timer-resolution constraint into the search space and directly optimizes the pulse sequence on realizable grid points, thereby avoiding the post-discretization approximation required by continuous-space methods.

To quantify the control capability over each selected component, this work utilizes a weighted total harmonic distortion (WTHD):
\begin{equation}
\label{equ16}
\rm{WTHD} = \frac{\sqrt{z_{n_{1}}^{2}+\ldots+z_{n_{N-1}}^{2}}}{\mathrm{MI}} \times 100\%,
\end{equation}
where \text{$z_{n}$} represents the amplitudes of the selected harmonics that are expected to be eliminated. Fig.~\ref{fig_7} (e) and (f) demonstrate the amplitudes of selected orders and the overall WTHD of the pulse sequence obtained by GCSA-PSO. As shown, the components are all controlled within a small range, and the fundamental deviations in the pulse sequence obtained by GCSA-PSO exhibit better control accuracy under practical digital-controller implementation.

To further confirm the effectiveness of the proposal, the cost landscape around the optimum is visualized in Fig.~\ref{fig_8}. To represent the four-dimensional data $(\vartheta_{1},\vartheta_{2},\vartheta_{3},J)$ in a three-dimensional coordinate system, color is used to indicate the cost value. Specifically, yellow corresponds to a lower cost value, whereas blue corresponds to a higher cost value. It can be observed that directly rounding the continuous-space optimum to the nearest timer grid point does not necessarily yield the global optimum in the discrete solution space. As shown in Fig.~\ref{fig_8} (b), under a timer frequency of 1~MHz, the discrete optimum may differ from the rounded continuous-space solution by up to two grid intervals. 

\subsection{Computational Analysis}

To further evaluate the computational efficiency of the proposed method, a comparative test is conducted among GCSA-PSO, PSO-NR, and classical PSO. For a fair comparison, the same initial conditions, maximum iteration number, population size, and algorithm parameters are adopted for all methods. Each case is independently repeated 100 times to reduce the influence of randomness, and the averaged results are shown in Fig.~\ref{fig_9}. As observed, PSO-NR requires the longest computational time among the compared methods. This is because the NR refinement is further applied after the PSO search to improve the solution accuracy, which inevitably introduces additional computational burden. Nevertheless, PSO-NR can achieve very high convergence accuracy in the continuous solution space, with the final cost value reduced to the order of $10^{-31}$ in the tested cases. However, such a continuous-space optimum does not necessarily correspond to the discrete-space optimum after timer-grid discretization. Consequently, the resulting pulse sequence may still suffer from implementation-induced optimality deterioration under practical digital-controller deployment, as shown in Fig.~\ref{fig_8}.

The dark-blue curve in Fig.~\ref{fig_9} represents the computational time of GCSA-PSO under different population sizes. It can be observed that GCSA-PSO requires less computational time than both classical PSO and PSO-NR under the tested conditions. This improvement mainly comes from the state-adaptive cost evaluation strategy. Unlike classical PSO and PSO-NR, which evaluates all particles using the full cost function, GCSA-PSO avoids full harmonic evaluation for particles that violate the sequence constraints or remain unchanged after grid adhesion. Moreover, the reduction in computational burden becomes more pronounced as the population size increases, since a larger population generally leads to more particles for which full cost evaluation is unnecessary.

\subsection{Implementation Verification under Various Scenarios}

In this subsection, the pulse sequences solved by GCSA-PSO under different modulation indices are organized into a LUT and deployed on the RT-Box3 controller. The test is conducted to evaluate the performance of the obtained HPPWM patterns under various operating conditions, including different modulation indices $\mathrm{MI}$, different numbers of switching angles $N$, and different fundamental frequencies $f_{\rm b}$.

Fig.~\ref{fig_10}(a) presents the voltage and current waveforms when the modulation index changes from 0.15 to 0.81 with $N=3$ under an RL load of $1~\Omega+10~\mathrm{mH}$. Fig.~\ref{fig_10}(b) illustrates the transition from three switching angles $(\mathrm{MI}=0.81, N=3)$ to five switching angles $(\mathrm{MI}=0.77, N=5)$. Fig.~\ref{fig_10}(c) shows the output waveforms when the fundamental frequency changes from $f_{\rm b}=50~\mathrm{Hz}$ to $f_{\rm b}=100~\mathrm{Hz}$. Due to the increased inductive reactance at higher frequencies, the output current decreases accordingly. Figs.~\ref{fig_10}(d)--(f) present the current spectra under the above operating conditions, while Figs.~\ref{fig_10}(g)--(i) compare the deviations of the controlled harmonic components between GCSA-PSO and the continuous-space solving method PSO-NR. It can be observed that, under the tested operating conditions, the pulse sequences obtained by GCSA-PSO achieve more accurate regulation of the selected harmonic components than those obtained by the conventional continuous-space solving method. This is because GCSA-PSO directly solves the pulse sequence in the practical discrete solution space, avoiding the additional approximation introduced by post-discretization.

\section{Conclusion}

 Harmonic programmed pulse width modulation (HPPWM) is a promising technique to tackle the harmonic issues for high-power systems operating at low switching frequencies. To achieve accurate solutions aligned with practical discrete digital controllers, this article proposes a Grid-Constrained State-Adaptive Particle Swarm Optimization (GCSA-PSO) strategy. By embedding the discrete timer constraints into the solution space and assigning state-adaptive cost evaluation strategy to particles, the proposed method directly obtains implementable pulse sequences while reducing unnecessary evaluations. Experimental results demonstrate that GCSA-PSO provides implementation-consistent HPPWM pulse sequences with improved computational efficiency and harmonic control accuracy under practical digital-controller implementation. The features are summarized as follows:
\begin{enumerate}
    \item The proposed method can be conveniently extended to different HPPWM formulations by modifying the corresponding cost function and constraint terms.
    \item The proposed method assigns different cost functions according to particle states, thereby avoiding unnecessary computational burden and offering shorter computational time.
    \item The proposed method can directly obtain the discrete pulse sequence without additional discretization, thereby preserving optimality in practical deployment.
    %The various evaluating strategies for different individuals prevents the inefficient assessment among the numerous particles. This can greatly improve the evaluation efficiency and reduce the computational time.
\end{enumerate}

Future work will focus on reducing the storage burden associated with HPPWM lookup tables.

\bibliographystyle{IEEEtran}
\bibliography{lib}

\balance

\end{document}